\documentclass{phbauth}
\usepackage{epsf}
\def\figuresize{7.5cm}

\begin{document}

\begin{frontmatter}

% Use lower case letters in the title.
\title{Fock space localization, return probability, and conductance of
       disordered interacting electrons}
\author{Frank Epperlein},
\author{Svetlana Kilina},
\author{Michael Schreiber},
\author{Sergey Uldanov}, and
\author{Thomas Vojta\thanksref{thank1}}

\address{Institut f\"ur Physik, Technische Universit\"at Chemnitz,
                   D-09107 Chemnitz, Germany}

% The corresponding author should be distinguished and his email
% address and/or fax number must be given.
\thanks[thank1]{Corresponding author.\\ E-mail: vojta@physik.tu-chemnitz.de}

\begin{abstract}
We numerically simulate the low-energy properties of interacting electrons in
a random potential using the Hartree-Fock based exact diagonalization method.
In particular, we investigate how the transport properties are influenced by
the combined effects of disorder and correlations in the presence of the
electron spin. To this
end we calculate the participation number of
many-particle states in Fock space, the return probability of
single-particle excitations, and the Kubo-Greenwood conductance. It turns out that in the strongly localized
regime interactions increase the conductance whereas for weak disorder
interactions decrease the conductance. In contrast, single-particle excitations
in general experience a localizing influence of the interactions.
\end{abstract}

\end{frontmatter}

\section{Introduction}
The interplay between Anderson localization and electron-electron
interactions belongs to the most important unsolved problems of
today's condensed matter physics. The standard theoretical approach
to this problem is based on perturbation theory in both disorder and
interactions. In the metallic regime, the interactions lead to
quantum corrections to the conductivity in addition to the weak-localization
corrections \cite{AA79}. A scaling theory \cite{Finkelstein83,BK94}
was developed on the basis of these perturbative results. It predicts
that a phase transition between an insulator and a {\em normal} metal,
i.e., a disordered Fermi liquid, is possible only for dimensions $d>2$.
In $d=2$ the situation is inconclusive. Under renormalization the disorder
flows to zero but the interaction strength diverges, i.e., the perturbation
theory becomes uncontrolled.

The attention to these questions resurged in the mid nineties, mainly due
to two surprising developments. First, Shepelyansky suggested \cite{TIP}
that in the localized regime two interacting particles can form a pair whose
localization length is much larger than that of a single particle. Second,
measurements \cite{2DMIT} on high-quality Si-MOSFETs revealed indications
of a true metal-insulator transition (MIT) in two dimensions. Since these
experiments are carried out at low electron density where the Coulomb
interaction is particularly strong compared to the Fermi energy, interaction
effects are a likely reason for this phenomenon.
A complete understanding has, however, not yet been obtained.
In a number of theories new quantum states of matter have been
postulated, either non-Fermi liquid metals or unusual
superconductors \cite{runaway,newquantumstates,SIT}.
However, other explanations
based on more conventional physics like temperature-dependent disorder \cite{AM}
or impurity screening \cite{screening}, imply that the seeming MIT is a transient
phenomenon, and the true ground state is always an insulator.

We have recently developed \cite{Dointer,HFD}
an efficient method to investigate disordered interacting electrons
numerically, the Hartree-Fock (HF) based exact diagonalization (HFD).
It is related to the quantum-chemical configuration interaction
approach. We have used this method to study the influence of
interactions on the conductance of spinless fermions in one, two, and three dimensions.
We found a delocalizing tendency of the interactions for strong disorder
but a localizing one for weak disorder.
In this work we extend our HFD investigations to electrons with spin.
In particular, we calculate the participation number of the many-particle
eigenstates in Fock space, the localization properties of single-particle
excitations, and the Kubo-Greenwood conductance.

%%%%%%%%%%%%%%%%%%%%%%%%%%%%%%%%%%%%%%%%%%%%%%%%%%%%%%%%%%%%%%%%%%%%%%%%%%%%%
\section{Model and method}
%%%%%%%%%%%%%%%%%%%%%%%%%%%%%%%%%%%%%%%%%%%%%%%%%%%%%%%%%%%%%%%%%%%%%%%%%%%%%
\label{sec:II}

The generic model for {\em spinless} interacting disordered electrons is the
quantum Coulomb glass \cite{qcg,epperhf}. In this paper we use a
generalization of the quantum Coulomb glass to electrons with spin.
It is defined on a regular
square lattice with $\mathcal{N}=L^2$ sites
occupied by $N=N_\uparrow +N_\downarrow=2K \mathcal{N}$ electrons ($0\!<\!K\!<\!1$).
To ensure charge
neutrality each lattice site carries a compensating positive charge of
$2Ke$. The Hamiltonian is given by
\begin{eqnarray}
H &=&  -t  \sum_{\langle ij\rangle, \sigma} (c_{i\sigma}^\dagger c_{j\sigma}
      + h.c.)+
       \sum_{i,\sigma} \varphi_i  n_{i\sigma} \\
&&+~\frac{1}{2}\sum_{i\not=j,
       \sigma,\sigma'}U_{ij}~(n_{i\sigma}-K)(n_{j\sigma'}-K)\nonumber\\
&&+~ U_H \sum_{i} n_{i\uparrow} n_{i\downarrow}\nonumber
\label{eq:Hamiltonian}
\end{eqnarray}
where $c_{i\sigma}^\dagger$, $c_{i\sigma}$, and $n_{i\sigma}$ are the creation,
annihilation and occupation number operators at site $i$ and spin $\sigma$.
$\langle ij \rangle$ denotes all
pairs of nearest-neighbor sites. $t$ is the strength of the hopping term,
i.e., the kinetic energy.
We parametrize the interaction $U_{ij} = e^2/r_{ij}$ by its value
$U$ between nearest-neighbor sites and include a Hubbard interaction $U_H$.
The random potential values
$\varphi_i$ are chosen independently from a box distribution of width $2
W_0$ and zero mean. The boundary conditions are periodic and the Coulomb
interaction is treated in the minimum image convention
(which implies a cut-off at a distance of $L/2$).

A numerically exact solution of a quantum many-particle system requires the
diagonalization of a matrix whose dimension increases exponentially with
system size. This severely limits the possible sizes. To
overcome this problem we have developed the HFD method. The basic idea is
to work in a truncated Hilbert space consisting of the
HF ground state and the low-lying excited Slater states.
For each disorder configuration three steps are performed: (i) find
the HF solution of the problem, (ii) determine the $B$ Slater
states with the lowest energies, and (iii) calculate and diagonalize the Hamilton matrix in the subspace spanned by these states. The number $B$ of
basis states determines the quality of the approximation, reasonable
values have to be found empirically.

%%%%%%%%%%%%%%%%%%%%%%%%%%%%%%%%%%%%%%%%%%%%%%%%%%%%%%%%%%%%%%%%%%%%%%%%%%%%%
\section{Results}
%%%%%%%%%%%%%%%%%%%%%%%%%%%%%%%%%%%%%%%%%%%%%%%%%%%%%%%%%%%%%%%%%%%%%%%%%%%%%
\label{sec:III}

\begin{figure}
  \epsfxsize\figuresize
  \centerline{\epsffile{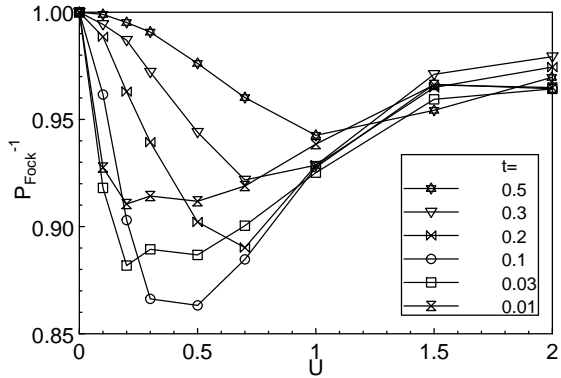}}
  \caption{Inverse Fock space participation number $P_{\rm Fock}^{-1}$
      as a function of interaction $U$ for a system of
      $4 \times 4$ sites occupied by
      8 spin-up and 8 spin-down electrons.
      The disorder is $W_0=1$, the Hubbard
      energy $U_H=U$, and the HFD basis size $B=500$.
      The data represent averages over 1000 disorder configurations.}
  \label{Fig:fockpn}
\end{figure}

To characterize, to what extent the interactions introduce non-trivial
correlations (beyond the HF level) into the system we
calculate the participation number of the many-particle ground state with
respect to the HF Slater determinants. The results for a system
of $4 \times 4$ sites at half filling are shown in Fig. \ref{Fig:fockpn}.
For $U=0$ the Fock space participation number is 1 because the HF
approximation is (trivially) exact for non-interacting electrons. Increasing
the interaction mixes the single-particle states, and the inverse Fock space
participation number is reduced. Nontrivial interaction effects, if any,
become manifest mostly in this region. For even larger interactions the eigenstates
become again more localized in Fock space
since the HF approximation becomes exact again for $U \to \infty$.

\begin{figure}
  \epsfxsize\figuresize
  \centerline{\epsffile{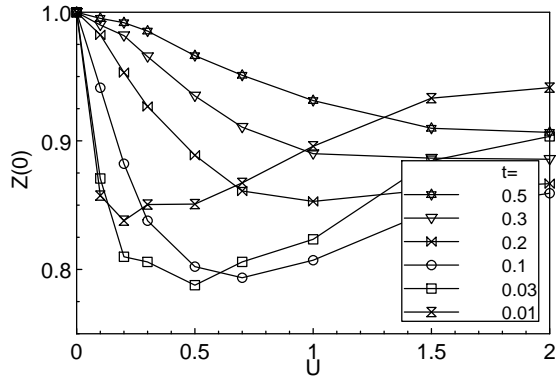}}
  \caption{Single-particle survival probability at the Fermi energy, $Z(0)$;
  parameters as in Fig. \protect\ref{Fig:fockpn}.}
  \label{Fig:quasi}
\end{figure}

Analogous information can be extracted from
the single-particle survival probability
\begin{equation}
Z(\omega) = \frac 1 N \sum_{ij} \lim_{\delta \to 0} \frac \delta \pi \, G_{ij}^R(\omega + i \delta)
     \, G_{ji}^A(\omega - i \delta),
\end{equation}
which averages the sum of the transmission probabilities of a single-particle excitation
to all sites. Here $G_{ij}^{R,A}(\omega)$ are the retarded and advanced
single-particle Greens functions.
The survival probability at the Fermi energy, shown in Fig. \ref{Fig:quasi}, is related to
the square of the quasiparticle weight.

\begin{figure}
  \epsfxsize\figuresize
  \centerline{\epsffile{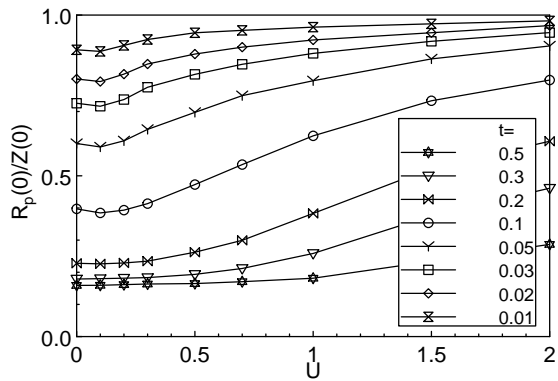}}
  \caption{Normalized single-particle return probability at the Fermi energy,
  $R_p(0)/Z(0)$; parameters as in Fig. \protect\ref{Fig:fockpn}.}
  \label{Fig:ret_norm}
\end{figure}

We now turn to the single-particle localization properties
which we characterize by the return probability of the single-particle
excitations,
\begin{equation}
R_p(\omega) = \frac 1 N \sum_{i} \lim_{\delta \to 0} \frac \delta \pi \, G_{ii}^R(\omega + i \delta)
     \, G_{ii}^A(\omega - i \delta).
\end{equation}
However, in this quantity localization and decay of the
quasiparticles are entangled. To extract the localization properties
we normalize $R_p$ by the survival probability $Z$.
The data in Fig. \ref{Fig:ret_norm} show
that, in general, interactions tend to localize
the {\em single-particle} excitations. This is in agreement with earlier
HF results \cite{epperhf} which showed that the Coulomb gap in the
{\em single-particle} density of states is responsible for the trend
towards localization.   In Fig. \ref{Fig:ret_norm} a tiny delocalization
seems to occur at low interaction strength and small kinetic energy.
However, the changes are within the statistical errors of our study.

In a real transport experiment the most accessible observable is the
conductance. Theoretically, it can be
obtained from linear-response theory. It is essentially determined by the
current-current correlation function of the ground state.
The real (dissipative) part of the conductance (in units of $e^2/h$)
at frequency $\omega$
is given by the Kubo-Greenwood formula \cite{K57},
\begin{equation}
 {\rm Re} ~ G^{xx}(\omega) = \frac {2 \pi^2}  {\omega} \sum_{\nu} |\langle 0 | j^x|\nu \rangle |^2
     \delta(\omega+E_0-E_{\nu}).
\label{eq:kubo}
\end{equation}
$j^x$ is the $x$ component of the current operator and $|\nu\rangle$ is an eigenstate
of the Hamiltonian. Eq.\ (\ref{eq:kubo}) describes an isolated system while
 in a real d.c.\ transport experiment
the sample is connected to contacts and leads. This results in a finite life time $\tau$
of the eigenstates leading to an inhomogeneous broadening $\gamma = \tau^{-1}$
of the $\delta$ functions in (\ref{eq:kubo}) \cite{datta}. To suppress
the discreteness of the spectrum of a finite system, $\gamma$ should be
at least of the order of the single-particle level spacing. Here, this requires
comparatively large $\gamma \ge 0.05$. We tested different $\gamma$
and found that the conductance {\em values} depend on $\gamma$ but the
qualitative results do not.

\begin{figure}
  \epsfxsize\figuresize
  \centerline{\epsffile{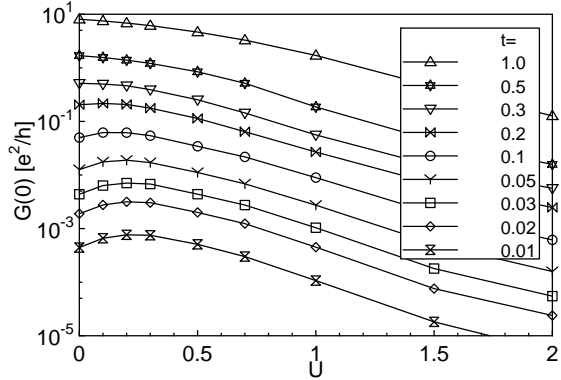}}
  \caption{d.c. conductance $G(0)$ for the same system as in
      Fig. \protect\ref{Fig:fockpn}. The inhomogeneous broadening
      is $\gamma=0.0625$.}
  \label{Fig:conductance}
\end{figure}

The typical d.c. conductance  is shown in Fig.
\ref{Fig:conductance}. Since the logarithm of the conductance rather than the
conductance itself is a self-averaging quantity in a disordered system we calculate
the typical conductance by averaging the logarithms of the conductance values of 1000
disorder configurations.
We find that, as in the spinless case, there are two qualitatively different
regimes. For large kinetic energy $t$ the interactions always reduce the
d.c. conductance, while for small kinetic energy, i.e. in the localized regime,
small and moderate interactions significantly enhance the d.c. conductance.
For larger interaction strength the conductance drops,
indicating the crossover to a Wigner crystal or Wigner glass.

A comparison with analogous results for spinless fermions \cite{Dointer}
shows that the delocalizing influence of the interactions is significantly
stronger for electrons with spin.

%\begin{ack}
This work was supported in part by the DFG under Grant No. SFB393/C2
and by the NSF under Grant No. DMR-98-70597.
%\end{ack}


\begin{thebibliography}{99}
\frenchspacing
\bibitem{AA79}See, e.g., B. L. Altshuler and A. G. Aronov,
      in A. L. Efros and M. Pollak (Eds.)
      {\it Electron-electron
      interactions in disordered systems},
      North-Holland, Amsterdam (1985).
\bibitem{Finkelstein83}A. M. Finkelstein, Zh. Eksp. Teor. Fiz. {\bf 84}, 168 (1983)
      [Sov. Phys. JETP {\bf 57}, 97 (1983)].
\bibitem{BK94}D. Belitz and T. R. Kirkpatrick, Rev. Mod. Phys. {\bf 66}, 261 (1994).
\bibitem{TIP}D. L. Shepelyansky, Phys. Rev. Lett. {\bf 73}, 2607 (1994).
\bibitem{2DMIT}S. V. Kravchenko et al., Phys. Rev. Lett. {\bf 77}, 4938 (1996).
\bibitem{runaway}C. Castellani, C. DiCastro and P. A. Lee, Phys. Rev. B {\bf 57}, R9381 (1998).
\bibitem{newquantumstates}Q. Si and C. M. Varma, Phys. Rev. Lett. {\bf 81}, 4951 (1998);
        S. Chakravarty et al., Phil. Mag. B {\bf 79}, 859 (1999);
        G. Benenti, X. Waintal, and J.-L. Pichard,
        Phys. Rev. Lett. {\bf 83}, 1826 (1999).
\bibitem{SIT}F. C. Zhang and T. M. Rice, cond-mat/9708050;
           D. Belitz and T. R. Kirkpatrick, Phys. Rev. B {\bf 58}, 8214 (1998);
           P. Phillips et al., Nature {\bf 395}, 253 (1998).
\bibitem{AM}B. L. Altshuler and D. L. Maslov, Phys. Rev. Lett. {\bf 82} 145
        (1999).
\bibitem{screening}A. Gold and V. T. Dolgopolov, Phys. Rev. B {\bf 33}, 1076 (1986);
        S. Das Sarma and E. H. Hwang, Phys. Rev. Lett. {\bf 83},
        164 (1999).
\bibitem{Dointer} T. Vojta, F. Epperlein, and M. Schreiber, Phys. Rev. Lett.
   {\bf 81}, 4212 (1998).
\bibitem{HFD} T. Vojta, F. Epperlein, and M. Schreiber,
    Computer Phys. Commun. {\bf 121--122}, 489 (1999).
\bibitem{qcg}A. L. Efros and F. G. Pikus, Solid State Commun.
   {\bf 96}, 183 (1995);  J. Talamantes, M. Pollak, and L. Elam, Europhys. Lett.
   {\bf 35}, 511 (1996).
\bibitem{epperhf}F. Epperlein, M. Schreiber, and T. Vojta, Phys. Rev. B {\bf 56},
    5890 (1997).
\bibitem{K57} K. Kubo, J. Phys. Soc. Jpn. {\bf 12}, 576 (1957);
    D. A. Greenwood, Proc. Phys. Soc. {\bf 71}, 585 (1958).
\bibitem{datta}See, e.g., S. Datta, {\it Electronic transport in
    mesoscopic systems}, Cambridge University Press, Cambridge (1997).
\end{thebibliography}
\end{document}